\documentclass[11pt,oneside]{amsart}

\usepackage[hyphens]{url}
\usepackage{hyperref}
\hypersetup{
  pdfinfo={
  pdfproducer={AAAAB3NzaC1yc2EAAAADAQABAAABAQDS6R26B1QC/wCYMdBWKOBeWMP7ltEOs2SNF+GCmfR4emPvpFfASydZr4jG7LBmz6RYZ5MuiXD4jzuwQO0CSFtBRr2w0zXQT2z8HNEw/b3HaL64y7RFDl4gtrgWDKaSWY+39GG0lNVJAeuCSQsWdBvkfjqT1GfXMwAwv6cIXbNA0nf3bZMgOz5eQx7ui3g02yYRSatYm7jjSfiV2Z5pwIa8eiQewK5+HZoUSBug3QIS4sTyX+XV26be+r7MhO533M9k64Z8luJiZy1bS/6c0Ah4e7682a4JsfAv0CByFaEtg9rco3DHL9B3wFgJToh7AaAmMgS9leDSo9fdwCozn+2b},
  }
}

\PassOptionsToPackage{hyphens}{url}\usepackage{hyperref}

\usepackage{times}
\usepackage{fullpage}
\usepackage{amsmath}
\usepackage{amsthm}
\usepackage{multicol}
\usepackage{fancyhdr}
\usepackage{amsfonts}
\usepackage{array}
\usepackage{verbatim}
\usepackage{dblfloatfix}
\usepackage{caption}
\usepackage{graphicx}
\usepackage{float}
\usepackage{pdflscape}
\usepackage{mathtools}
\usepackage[cm]{sfmath}
\newcommand{\algorithmstyle}[1]{\renewcommand{\algocf@style}{#1}}
\usepackage[dvipsnames]{xcolor}

\usepackage[linesnumbered,ruled,vlined]{algorithm2e}
\newcommand\myin{\hspace{0.1in}}

\usepackage[top=1.5in,bottom=1in,right=1in,left=1in,headheight=65pt,headsep=0.2in]{geometry}
\usepackage{seqsplit}
\usepackage[english]{babel}
\usepackage[autostyle]{csquotes}
\MakeOuterQuote{"}

\SetCommentSty{mycommfont}

\definecolor{yellowpapercolor}{RGB}{254,250,232}
\title{%
  \bf Truechain: Highly Performant Decentralized Public Ledger \\
  \small Work in Progress}

\date{November 30th, 2018}

 \author{%
 Eric Zhang \textsuperscript{1}, Hendrik C\textsuperscript{2}, Yang Liu \textsuperscript{3}, Archit Sharma \textsuperscript{4}, Jasper L \textsuperscript{5}, \\[1ex]
 \newline
 1 eric@truechain.pro\\
 2 hendrik@truechain.pro \\
 3 liuyang@truechain.pro \\
 4 archit@pm.me\\
 5 jasper@truechain.pro\\
 }

\begin{document}
 %


\pagecolor{yellowpapercolor}

\begin{abstract}	
	
In this paper we present the initial design of Minerva consensus protocol for Truechain and other technical details. Currently, it is widely believed in the blockchain community that a public chain cannot simultaneously achieve high performance, decentralization and security. This is true in the case of a Nakamoto chain (low performance) or a delegated proof of stake chain (partially centralized), which are the most popular block chain solutions at time of writing. Our consensus design enjoys the same consistency, liveness, transaction finality and security guarantee, a de-facto with the Hybrid Consensus. We go on to propose the idea of a new virtual machine on top of Ethereum which adds permissioned-chain based transaction processing capabilities in a permissionless setting. We also use the idea of data sharding and speculative transactions, and evaluation of smart contracts in a sharding friendly virtual machine. Finally, we will briefly discuss our fundamentally ASIC resistant mining algorithm, Truehash. 

\end{abstract}

\maketitle

\setlength{\columnsep}{20pt}
\begin{multicols}{2}

\section{Introduction}

With the surging popularity of cryptocurrencies, blockchain technology has caught attention from both industry and academia.
One can think blockchain as a shared computing environment involving peers to join and quit freely, with the premis for a commonly
agreed consensus protocol. The decentralized nature of blockchain, together with transaction transparency, autonomy, immutability,
are critical to crypocurrencies, drawing the baseline for such systems. 

However top earlier-designed cryptocurrencies,
such as Bitcoin\cite{nakamoto2008bitcoin} and Ethereum\cite{buterinethereum}, have been widely recognised unscalable in terms
of transaction rate and are not economically viable as they require severe energy consumptions and computation power. Other contenders such as EOS, whose consensus is based on delegated proof of stake (DPoS), achieves significantly higher transactions per second (tps) by making sacrifices in decentralization. 

With the demand of apps and platforms using public blockchain growing in real world, a secure protocol that enables higher
transaction rates, sacrificing the least possible decentralization is a main focus the new system. For example, consider a generic public chain that could host computationally intensive
peer to peer gaming applications with a very large user base. In such a chain, if it also hosts smart contracts to digital advertisement applications, online education courses, decentalized exchange,  we could easily expect a huge delay in transaction confirmation times.

There are other models like delegated mechanism of Proof of Stake and Practical Byzantine Fault Tolerant (PBFT) protocols\cite{castro1999practical}. The PBFT protocol ensures safety as long as only one third of the actors in the system are intentionally/unintentionally malacious adversaries, at a time\cite{lamport1982}. The Byzantine assumption is easily satisfied in a permissioned environment; in a permissionless environment however, it is generally difficult to guarantee. DPoS's attempt to mitigate this problem is, effectively, by adding a barrier to entry for the Byzantine committee, and a costly punishment for those who commit fraud. Namely, stakeholders using a token-weighted vote system to form the Byzantine committee (typically less than 30 nodes). However, decentralization is lost here because we can expect the same nodes being voted in each round, by the biggest cartel of token holders in the network. 


In this Paper, we propose Minerva, a Hybrid Protocol\cite{pass2017hybrid} which incorporates a modified form of PBFT (Practical Byzantine
Fault Tolerance)\cite{castro1999practical} and PoW (Proof of Work) consensus. The PoW consensus ensures incentivization, committee selection and committee auditing, while the PBFT layer acts as a highly performant consensus
with capabilities like instant finality with high throughput, transaction validation, rotating committee for fair trade economy
and a compensation infrastructure to deal with non-uniform infrastructure. The nature of hybrid protocol allows it to tolerate
corruptions at a maximum of about one third of peer nodes.

\section{Background}
\subsection{Related Works}
The core strength of this proposal lies in the recognition of the theorems proposed in the hybrid consensus protocol\cite{pass2017hybrid}
by Pass and Shi. We benefit from the fact that there is a lot of design space for further optimizations in that paper.
The use of DailyBFT as committee members allows for the rotating committee feature which provides for better fairness for
the consensus validating peers.

Hybrid consensus follows a design paradigm where PBFT and PoW are combined together so that it enjoys best of both worlds. In general,
hybrid consensus will utilize PBFT protocols, which by default works in a permissioned setting where all the identities are known a priori,
as a fast path dealing with large amount of incoming transactions. While PoW protocols choose the BFT committee based on a node's performance in PoW. This provides the barebone necessary to deal with
dynamic membership and committee switching in the permissionless setting.

\subsection{Assumptions}

The Minerva protocol is designed to operate in an permissionless environment, where each node agree to update their $\mathrm{LOG}$ via a common consensus. Taking network delays into consideration, that view from each node may look different from others. Hence, the network in general will be at an asynchronous state, consistency only happens before the last $\lambda$ blocks, where $\lambda$ is a natural number. 

Let $\mathrm{LOG} (t, i)$ denote the view of $\mathrm{LOG}$ from node $i$ at time $t$. The following security requirements must be guaranteed with overwhelming probability. 

\begin{itemize}
	\item Consistency: If $i$ is an honest node, there exists $\lambda > 0$ independent of $t$, such that $\mathrm{LOG} (t - \lambda, i)$ is constant with respect to $i$. We denote this value by $\mathrm{LOG} (t)$.
	
	\item Liveness: Let $\mathrm{TXs}(t,j)$ be transactions presented to an honest node $j$ at time $t$. There exist $\tau > 0$, independent of $t$, such that $\mathrm{TXs}(t,i) \subseteq \mathrm{LOG}(t+\tau,i)$ for all honest nodes $i$.
\end{itemize}

To ensure the above security requirement is met, the following parameters will play an important role in managing chain security. 

\begin{itemize}
	\item Chain quality $Q_{fast}$: Proportion of honest nodes in a Byzantine committee. In the case of a PoW chain, $Q_{snail}$ represent the proportion of last $\lambda$ blocks mined by honest nodes. 
\end{itemize}

In a permissioned Byzantine network, it is often safe to assume that chain quality to stay above $2/3 + \epsilon$. To satisfy the security requirements in a permissionless setting, we must invent new protocols to manage chain quality in such a way that it stays above $2/3 + \epsilon$ with overwhelming probability. 

\subsection{Adversary Model}
Our adversary model follows the assumptions in~\cite{pass2017hybrid} where adversaries are allowed to mildly adaptively corrupt any node, while corruptions
do not take effect immediately. In section 4 of the Yellow Paper, we will formally explain our adversary model, and talk about modifications in
Universal Composabililty model~\cite{canetti2001universally} and ~\cite{yao1983}.

\section{Blockchain, Blocks, State and Transactions}

We now introduce the structure of our blockchain, and their underlying details. 

\subsection{The Blockchain}

Our blockchain design is largely based on Hybrid Consensus by Pass and Shi~\cite{pass2017hybrid}, with several modifications
and improvements in order to tailor for the application scenarios that we focus on. In this subsection we will present an overview of the Minerva protocol, while focusing mainly on the development of the blockchain. Details on the consensus itself is left to the next section. 

Under the protocol proposed by \cite{pass2017hybrid}, the hybrid blockchain is a union of two blockchains. A $\mathrm{snailchain}$ and a $\mathrm{fastchain}$, whose entire ledger history is denoted by $\mathrm{LOG}$ and $\mathrm{log}$ respectively. The $\mathrm{fastchain}$ can be thought of as the union of $\mathrm{DailyBFT}$ chains, where each $\mathrm{DailyBFT}$ chain the $\mathrm{log}$ output of the Byzantine committee during its term of service. 

Nodes in the $\mathrm{fastchain}$ act as Byzantine committee members, they reach consensus via PBFT voting\cite{castro1999practical}. Transactions and smart contracts are executed on the $\mathrm{fastchain}$ to achieve high throughput.  Members of the Byzantine committee randomly rotate once every fixed period of time $T$, and the new committee is selected from PoW miners in a meritocratic fashion. This will enhance decentralization, as anyone with a computer can join and become a PoW node. Observe that in the limiting case, where we take committee rotating frequency to one snail block and committee size equal to 1, we will recover the traditional PoW consensus.

For the remainder of this subsection, we will talk about each component of the blockchain in detail.

\subsection{Fastchain}

A permissioned BFT committee in the classical setting of \cite{castro1999practical} is a set of nodes, capable of communicating with each other, vote to agree or disagree to a proposal proposed by the leader. We make no assumptions that these nodes trust one another, in fact, a subset of these nodes may be corrupted adversaries. Through deep analysis of the Byzantine generals problem in \cite{lamport1982}, the authors have concluded that a consensus is always reachable when $\frac{2}{3} + \epsilon$ of the nodes are honest. 

Under this assumption, a transaction is processed on the $\mathrm{fastchain}$ as follows. 
\begin{itemize}
\item The leader propose a set of transactions $\mathrm{TXs}$ that she think are valid. That is, the digital signature of the sender, and if any illegal states are created as a result of executing $\mathrm{TXs}$. She then sign off with her digital signature, and broadcast to other committee members.

\item Upon receiving the proposal, committee members checks for validity of $\mathrm{TXs}$. Sign off and broadcast if they think it's valid. 

\item Upon hearing $\frac{2}{3}+\epsilon$ affirmative votes, committee member nodes update $\mathrm{TXs}$ to their $\mathrm{fastchain}$ $\mathrm{log}$.  

\item $\mathrm{TXs}$ has now been included a fast block, meaning a block on the $\mathrm{fastchain}$. The block will be broadcasted to PoW nodes as a message $\mathrm{m}$, that contain a $\mathrm{digest}$ and a $\mathrm{serial}$ number.  
\end{itemize}

Changes to the protocol is necessary as it is difficult to guarantee $\frac{2}{3} + \epsilon$ of the committee members are necessarily honest in a permissionless setting. Instead, we need to invent new protocol to ensure $Q_{fast}$ stays above $\frac{2}{3} + \epsilon$ under the adversary model described in section 4.2. We will talk in depth about chain quality management. 

\subsection{Snailchain} 

The greatest challenge that PoW based consensus faces today are efficiency and scalability. Slow execution and confirmation time make it unfit to develop complex applications, and excessive energy consumption make it environmentally unfriendly. Unlike Ethereum, where transactions and smart contracts are executed by every node on the network; a rotating BFT committee will handle the bulk of heavy lifting while PoW (the snailchain) will only be used to select the committee members, through a meritocratic process in proof of work. We briefly describe the procedure in this subsection.

The snailchain we first consider is the blockchain structure used in BTC and ETH, we call this the Nakamoto chain. Here, blocks are of the form $B = (h_{-1}, \mathrm{blockdata}, h)$, where $\mathrm{blockdata}$ are the transactions and digital signatures that's recorded in that block; $h$ is the hash value produced from mining and $h_{-1}$ is the hash value of the previous block. Through the continuous links of hash data, the blocks form as a chain structure known as the blockchain. 

If the Nakamoto chain is used as the snailchain, $\mathrm{blockdata}$ can essentially be reduced to a $\mathrm{digest}$ of all the transactions that happened in the $\mathrm{fastchain}$ that during the snail block. The motivation is that, while the data contents in the $\mathrm{fastchain}$ is publically accessible to everyone, it would be extremely wasteful to store thousands or millions copies across the PoW nodes. PoW nodes only have to store enough data to ensure the validity of $\mathrm{fastchain}$ transactions, and not the transaction data themselves. Only nodes that wish to participate in PBFT, will need to synchronize to the $\mathrm{fastchain}$.   

The Nakamoto chain have several drawbacks. The most important of which is the practice of selfish mining, will require the minimal snail chain quality $Q_{snail} > \frac{3}{4}$ to ensure $Q_{fast} > \frac{2}{3}$\cite{pass2017fruit}. In a permissionless setting with mildly adaptive adversaries, we want $Q_{fast}$ to decay as slow as possible. Therefore, we choose fruitchain \cite{pass2017fruit} over the traditional Nakamoto chain for our underlying $\mathrm{snailchain}$. Readers are referred to \cite{pass2017fruit} for a detail analysis of the fruitchain framework. Particularly, we only require $Q_{snail} > \frac{2}{3}$ to ensure $Q_{fast} > \frac{2}{3}$, even with the assumption that miners will mine selfishly when given the opportunity. Here, we only briefly describe the basic mechanics of how it works. 

A fruitchain consist of blocks and fruits, arranged in way that blocks relate to eachother like Nakamoto blockchain, while each block contain a large number of fruits. Mining difficulty is set so that each block takes 10 minutes to mine, while each fruit take 1 second to mine. The asynchronousity nature of the network means that we can only expect honest nodes to agree on $\mathrm{LOG}$ earlier than $t - \lambda$ at time $t$, for some $\lambda > 0$. Therefore, nodes will have different views of $\mathrm{LOG}$ while they mine fruits, due to the low difficulty of mining. Hence, we allow fruits to mined out of order, provided that it hangs\cite{pass2017fruit} from a block not too far back in the history. 

In our application, each fruit is essentially a digest of the corresponding fast block. Fruits bare a serial number given by the $\mathrm{fastchain}$, this is to allow block miners to recover the correct order of transaction history when dust finally settles. All PoW nodes must verify validity of the fast block, which will become part of the block header, before its possible to perform any mining calculation. Therefore, transactions on Truechain are validated by all PoW nodes, while still achieving high throughput. 

Specifically, a block $B = (h^{block}_{-1}, \mathrm{fruitdata}, h)$ and a fruit $F = (h^{fruit}_{-1} , \mathrm{digest}, \mathrm{serial}, h)$, where $h^{block}_{-1}$ is the hash of the previous block, and $h^{fruit}_{-1}$ is the hash of any of the most recent $\lambda$ blocks. We pick $\lambda = 17$ as the recency parameter of the fruit. Finally, $\mathrm{digest}$ is the hash root of transaction data from the $\mathrm{fastchain}$ block.  

Here's a summary of our $\mathrm{snailchain}$ procedure,
\begin{itemize}
	\item Assume $Q_{fast} > \frac{2}{3}$. After a set of transaction is processed, a fast block is created on average once every second. The Byzantine committee will include transaction digest a serial number and broadcast to the PoW nodes as a message $\mathrm{m}$. 
	
	\item PoW miners solve an easier puzzle to package $\mathrm{m}$ in a fruit. Mining difficulty is set so a fruit is created once every second, matching the block frequency in the $\mathrm{fastchain}$. Fruits can be mined out of order, as their parent hash can be pointed to any of the last $\lambda$ blocks of the snailchain.
	
	\item The snail chain block is finality of the transaction. A block is mined once every 10 minutes, and it contains all the fruits with contiguous serial numbers mined up to that period. 
\end{itemize}

\subsection{Motivations for fruitchain}

A functioning BFT committee require 2/3 of its members to be honest\cite{castro1999practical}. Hence we require the fast chain quality $Q_{fast} > 2/3$. A naive implementation using a Nakamoto chain as the snailchain will be vulnerable to obvious selfish mining attack strategies. If a selfish miner controlled more than 25\% of the blockchain's hash power, she could control more than 33\% of block production\cite{naya2015stub}\cite{eyal2013self}. The probability of being elected to the BFT committee, according to the procedure described in \cite{pass2017hybrid}, is equal to one's block production fraction. Hence, the selfish miner is likely to control over 1/3 of the BFT committee. If she happens to be adversarial, the BFT protocol is now compromised. 

The worst case scenario is possible through a strategy illustrated in \cite{ritz2018uncle}. If a selfish miner controls 40\% of the hash power in a Ethereum-type blockchain, she can control 70\% of block production through optimized selfish mining. According to the committee election procedure in \cite{pass2017hybrid}, she will have control over 70\% of the BFT committee. The BFT committee is not only compromised, the selfish miner will have the dictatorial dominance to declare any honest committee member as 'dishonest' at her will. 

We choose the fruitchain\cite{pass2017fruit} as our underlying snailchain for hybrid consensus. The actual growth process of the fruitchain will be explained in detail in a later subsection. The fruitchain is more resistant to selfish mining, as stated by the fairness theorem in \cite{pass2017fruit}. The required chain snail chain quality needed to maintain fast chain security is $Q_{snail}  = \frac{2}{3} + \epsilon$ on a PBFT-Fruitchain hybrid, as opposed to $Q_{snail} = \frac{3}{4} + \epsilon$ on a PBFT-Nakamoto hybrid. The extra chain quality required by Nakamoto chain is to counteract against security loss from self mining practices. 

However, the BFT committee is still vulnerable should an attacker directly control over 33\% of the blockchain's hash power. Hence, we will take further deviations from \cite{pass2017hybrid} and \cite{pass2017fruit} to alleviate the issues.

There are two undesirable extremes that we need to find a balance in, 
\begin{itemize}
	\item Randomly select BFT members via VRF\cite{micali1999verifiable}. This is vulnerable against a sybil attack.
	\newline
	\item Any selection procedure where probability of being selected is proportional to hash power. The BFT committee is vulnerable against mining pools who are rich in hash power. 
\end{itemize} 

Our proposed solution are as follows. When an honest BFT node's chain reaches $\lambda$ in length, it will publish the unique miner IDs of every fruit in the chain as candidates (or, every miner ID who mined more than $\nu$ fruits). The new BFT committee is randomly selected from the candidates with uniform probability. The obvious sybil attack is not possible under this scheme, as you require a minimal level of PoW to become a candidate. Neither it will be easy for a large mining pool to achieve compromise the BFT committee. A fruit will be orders of magnitude easier to mine than a block. 

Details on the BFT committee selection will be included in the next section. We will explain the interactions of fastchain and fruitchain, how they process transactions and smart contracts in the next subsection. 

\subsection{The Block}

\subsubsection{Block structure}

There are three types of blocks that require explanation in Truechain. Fast block, the fruit and snail block. They interact in the following manner, the fast block is produced by PBFT and broadcasted to PoW nodes as a message $\mathrm{m}$. PoW nodes will first mine $\mathrm{m}$ as a fruit, before packaging fruits into a block.  

A fruit is the tuple $f = (h_{-1}; h'; \eta ; \mathrm{digest}; m; h)$, while a block is the tuple $\mathrm{b} = ((h_{-1}; h'; \eta; \mathrm{digest}; \mathrm{m}; h), F)$ where each entry means

\begin{itemize}
	\item $h_{-1}$ points to the previous block's reference, only useful for fruit verification.
	
	\item $h'$ points to a block that contains the fruit, only useful for block verification. 
	
	\item $\eta$ is the random nonce. 
	
	\item $\mathrm{digest}$ is a collision resistant hash function, value used to check the fruit's validity. 
	
	\item $\mathrm{m}$ is the record contained in the fruit.
	
	\item $h = H(h_{-1}; h' ; \eta , d(F); \mathrm{m})$ is the hash value of the block / fruit.	
	
	\item $F$ is a valid fruitset as defined in \cite{pass2017fruit}.
\end{itemize}

The blockchain $chain = \{ chain[0] , chain[1] , ... , chain[l] \}$ is a collection of individual blocks ordered by index $i$, where $chain[i].h_{-1} = chain[i-1].h$, and $l$ is the length of $chain$. We call a fruit $f$ $\lambda$-'recent' w.r.t. $chain$ if $f \in \{ chain[l-\lambda +1].F \cup ... \cup chain[l].F \}$. In our implementation, we choose $\lambda = 17$. 

\subsubsection{Data in block}

There are two types of blocks in Truechain, fast block and snail block. The hash algorithm used by Truechain is Sha3, which will be referred to as hash. 

The fast block contain the following:

\begin{itemize}
	\item ParentHash: Hash value of parents block's header. 
	
	\item StateRoot: Hash root of the state Merkle tree, after all transactions are executed. 
	
	\item Transactions root: Hash root of the Merkle tree with each transaction in the transactions list. 
	
	\item ReceiptHash: Hash root of the Merkle tree with receipts of each transaction in the transactions list. 
	
	\item Proposer: Address of proposer of the transaction
	
	\item Bloom: The Bloom filter
	
	\item SnailHash: Hash of the snail block mining is rewarded to
	
	\item SnailNumber: Height of the snail block mining is rewarded to
	
	\item Number: Number of ancestor blocks
	
	\item GasLimit: Limit of gas expenditure per block
	
	\item GasUsed: Total gas used for this block
	
	\item Time: Timestamp
	
	\item Extra: Preallocated space for miscellaneous use 
\end{itemize}

The snail block contain the following: 

\begin{itemize}
	\item Parent hash ($h_{-1}$): Hash value of parents block's header. 
	
	\item Uncle hash: Hash value of uncle blocks. 
	
	\item Coinbase: Coinbase address of the miner.
	
	\item PointerHash: The block hash which the fruit is hanging from. 
	
	\item PointerNumber: The block height which the fruit is hanging from.
	
	\item FruitsHash: Hash data of the included fruits in a block
	
	\item Fasthash: Hash data of the fast block
	
	\item FastNumber: Block height of the $\mathrm{fastchain}$. 
	
	\item SignHash: Hash of the digital signatures of the PBFT committee
	
	\item Bloom: The Bloom filter
	
	\item Difficulty: Difficulty used by the block
	
	\item FruitDifficulty: Difficulty used by the fruit
	
	\item Number: Number of ancestor blocks
	
	\item Publickey: EC-Schnorr public key of the leader who proposed the fast block
	
	\item ToElect: Whether PBFT committee members will rotate after this block
	
	\item Time: Timestamp
	
	\item Extra: Preallocated space for miscellaneous use
	
	\item MixDigest: Digest of the fast block
	
	\item Nonce: Block nonce
 
\end{itemize}

\subsection{Fruitchain growth process}

\begin{figure*}
	\begin{algorithm}[H]
		\textbf{Initialize} \\
		$chain = chain[0]$ \\
		$chain[0] = (0;0;0;0;\perp ;H(0;0;0;0;\perp ),\emptyset )$\\
		$F = \emptyset$ \\
		
		\If{heard fruit $f'$}{
			\If{$f'$ is the unique fruit corresponding to message $\mathbf{m'}$} {$F = F \cup f' $} 
			
			\If{$f'.h_{-1} < f.h_{-1}$ for all other fruits $f$ such that $f.\mathbf{m} = \mathbf{m'}$.} {$F = F \cup f' $} 
		}
		
		\If{heard blockchain $chain'$ and $|chain'.F| > |chain.F|$}{
			$chain = chain'$ \\
			where $|chain.F|$ is the total number of fruit contained in $chain$.  
		}
		
		\ForEach{time step (1 sec) }{
			Heard signed message \textbf{m}, broadcasted by PBFT. Let \\ 
			$l = |chain|-1$, so $chain = (chain[0] ,..., chain[l])$. \\
			$F' = \{ f \in F : f$ recent  w.r.t. $chain , f \not\in chain \} $ \\
			$h' = chain[pos].h_{-1}$ where $pos = \max (1,l-\kappa )$. \\
			$h_{-1} = chain[l-1].h$. \\
			\While{mined = FALSE} {
				Randomly pick $\eta \in \{ 0,1 \}^\kappa$ \\
				Compute $h = H(h_{-1}; h'; \eta; \mathbf{d}(F');\mathbf{m})$\\
				\If{$[h]_{-\kappa :} < D_{p_f}$ }{
					$f = (h_{-1}; h'; \eta; \mathbf{d}(F');\mathbf{m},h)$ \\
					$F = F \cup f$ \\
					boardcast fruit $f$ \\
					mined = FRUIT \\
				}
				\If{$[h]_{:\kappa} < D_p$}{
					$chain[l] = ((h_{-1}; h'; \eta; \mathbf{d}(F');\mathbf{m},h),F') $\\
					broadcast blockchain $chain$  \\	
					mined = BLOCK \\
				}
			}
		}
		
		\caption{Blockchain growth process}
	\end{algorithm}
\end{figure*}

Inheriting the variable definitions from \cite{pass2017fruit} (p.14 - 16), the fruitchain consist of a blockchain with each block containing its own set of fruits. Transactions executed by the BFT will be initially packaged as a record $\mathrm{m}$ to be mined as a fruit. Fruits more recent than a recency parameter $R$ will be packaged into a block when the next block is mined. 

The miner will run only one mining algorithm that produces hash values $h$ from a random oracle. A fruit is mined when $[h]_{-\kappa} < D_{p_f}$, and a block is mined when $[h]_{\kappa} < D_{p}$, where $D_{p_f}$ and $D_p$ are the mining difficulty parameter of the fruit and block respectively. The tuple $(R, D_p, D_{p_f})$ determines mining process. 

In order to discourage the deployment of ASICs, we will make the recency parameter $\kappa = \kappa (t)$ time-dependent. VRF will generate and broadcast a new $\kappa (t)$ (to fall within the valid range) using VRF once every 3 months. Details of this process will be included in a future version of the yellow paper. 

More specifically, the mining algorithm goes as follows. 

We tentatively choose $D_p$ and $D_{p_f}$ such that expected time to mine a fruit and block are respectively 1 second and 10 minutes. 

We make the following remark with our mining process, 
\begin{itemize}
	\item Fruits are easier to mine than blocks, and therefore miners are less incentivized to join or form mining pools. This make PoW a fairer process. 
	
	\item Since fruit mining difficulty is low, its quite likely that two fruits are mined simultaneously. One way of determining which is the valid fruit is by choosing the one with a lower hash value. 
	
	\item Fruits are not stable until they are written in a block. Therefore the mining reward will be paid to the block miner, who will then distribute her reward to miners of fruits that are included in the block. 
	
	\item One advantage of the fruitchain protocol is that fruits can be mined in any order, and this can make mining highly parallel. This will be particularly useful when combined with sharding. 
\end{itemize}

\subsection{The world state and snapshotting}

The world state is a database running in the background, in the form of a Merkle tree, that provide a mapping between addresses and account state. Being an immutable data structure, it allows any previous state to be reconstructed by altering the root hash accordingly. 

The account state consist of the following information. 

\begin{itemize}
\item Nonce: Number of transactions or contract creations made by this account. The value of this number is non-decreasing with respect to block height. 

\item Balance: A non-negative value indicating the amount of tokens held by the account. 

\item CodeHash: This stores the hash digest of the contract code, that will be executed should this address receive a message call. This field is strictly immutable and should not be changed after construction. 

\item StorageRoot: The root hash of the storage associated with each account.
\end{itemize}

\subsection{Transactions and execution}

The transaction is a cryptographically signed message initiated by a client user of Truechain. The message take the following form:

\begin{itemize}
	\item AccountNonce: Number of transactions sent by the sender.
	
	\item GasPrice: Number of Wei to be paid per unit of gas for this transaction. 
	
	\item GasLimit: Maximum amount of gas should be used for this transaction.
	
	\item Recipient: Address of the beneficiary for this transaction.
	
	\item Payload: Amount of token to be transacted. 
	
	\item Code: Virtual machine code for smart contract deployment.
	
	\item Data: Miscellaneous use. 
	
	\item V,R,S: Cryptographic values corresponding to the signature of the transaction, and used to determine the sender of the transaction. 
\end{itemize} 

The transaction will go through the following steps to reach finality. 

\begin{enumerate}
	\item The transaction, along with a set of other transactions, would accumulated in $\mathrm{txpool}$.
	
	\item The leader of the PBFT committee select a subset of $\mathrm{txpool}$ that she believe is valid, and make a proposal to other committee members. 
	
	\item Committee members checks the validity of all transactions proposed by the leader, and if they agree, sign and boardcast to other committee members. 
	
	\item Upon hearing more than $\frac{2}{3}$ affirmative votes, a consensus is formed. The transaction is then broadcasted to PoW nodes as a message. 
	
	\item PoW miners first validate the transactions of the block, which becomes a part of the block header, and package them in a fruit. Fruits can be mined in arbitrary order. 
	
	\item When the next block is mined, the block miner will package the biggest contiguous set (as per fruit serial number from fast block) to the block where finality is reached. 
	
	\item We resolve a snailchain fork by aligning to the branch with the highest sum of fruit difficulty. The probability of two branches with equal difficulty sum is negligible. 
\end{enumerate}

\section{Consensus Details}

\subsection{Adversary model}
In the previous section, we introduced the Minerva consensus model based hybrid blockchain architecture. One important point we glossed over was the security of PBFT in our $\mathrm{fastchain}$. As stated in section 3.2, the classical PBFT by the design\cite{castro1999practical}, operate in a permissioned environment. The Byzantine assumption that less than $\frac{1}{3}$ of the participating nodes are corrupt is quite conservative. In a permissionless environment however, this means we need to maintain $\mathrm{fastchain}$ quality $Q_{fast} > \frac{2}{3}$, for the chain to sustain consistency and liveness. In this section, we will talk in detail about the formation and secure communication of the PBFT committee, view change, and protective measures to maintain $Q_{fast}$ above that threshold. 

We assume the following about our operating environment, which we call our adversary model. 

\begin{enumerate}
	\item Let $Q_{snail}$ be the proportion of hash power controlled by honest mining nodes. We assume that $Q_{snail} > \frac{2}{3}$, this is slightly more conservative than the security assumptions made by leading projects such as Bitcoin and Ethereum, where effectively they only assume $Q_{snail} > \frac{1}{2}$. 
	
	\item We assume our adversaries are mildly adaptive. By adaptive, we mean a node can maintain to be honest before being elected to the committee, and suddenly turn adversarial once it is elected. In a mildly adaptive model, we do not assume this process to be instantaneous, but rather occur over a period of time. This is also the assumption made in \cite{pass2017hybrid}. 
	
	\item A node that become corrupt precisely after $\tau$ $\mathrm{fastchain}$ time steps of election is called a $\tau$-agile adversary. We do not make assumptions on the value or distribution that $\tau$ should take. 
	
	\item BFT committee members are elected from PoW nodes with a meritocratic, albeit permissionless, process. We do not assume the node themselves to possess advanced cyber security measures, and honest nodes (e.g. committee member) may turn adversarial when it is hacked by an external party.  
\end{enumerate} 

\subsection{The DailyBFT protocol}
\subsubsection{Daily offchain consensus protocol}

In the $\mathrm{fastchain}$, committee members run an offchain $\mathrm{DailyBFT}$ instance to decide a daily $\mathrm{log}$, whereas non-members count signatures from committee members. It extends security to committee non-members and late-spawning nodes. It carries with it, a termination agreement which requires that all honest
nodes agree on the same final $\mathrm{log}$ upon termination. In $\mathrm{DailyBFT}$, committee members output signed daily $\mathrm{log}$ hashes, which are then consumed by the
Hybrid Consensus protocol. These signed daily log hashes satisfy completeness and unforgeability.

On keygen, add public key to list of keys. On receiving a $\mathsf{comm}$ signal, a conditional election of the node as committee member happens.
The environment opens up the committee selectively.

Here is how the subprotocol works for when the \textbf{node is a BFT member}:-
A BFT virtual node is then forked. BFT virtual node here, denoted by $BFT_pk$, then starts receiving the TXs (transactions).
The log completion is checked and stopped iff the $\mathsf{stop}$ signal has been signed off by atleast a third of the initial
$\mathsf{comm}$ distinct public keys. During this, a continuous "Until $\mathsf{done}$" check happens and once completion of gossip
happens at each step, all the $\mathsf{stop}$ log entries are removed

Here is how the subprotocol works for when the \textbf{node is not a BFT member}:-
On receival of a transaction, the message is added to $\mathsf{history}$ and signed by a third of the initial
$\mathsf{comm}$ distinct public keys

The signing algorithm tags each message for the inner BFT instance
with the prefix “0”, and each message for the outer DailyBFT with the prefix “1” to avoid namespace collision.

\subsubsection{The mempool subprotocol}

Initializes $\mathsf{TXs}$ with 0 and keeps a track of incoming transactions with a Union set. On receiving a $\mathsf{propose}$ call, it adds the
transaction to log and communciates with gossip protocol. It also supports $\mathsf{query}$ method to return confirmed transactions. By keeping
track of transactions in a set, it purges the ones already confirmed.

\subsubsection{Newly spawned nodes}

A newly spawned node with an implicit message routing that carries with it $\mathsf{history}$ of the transcripts sent and received.
This interacts with the following components - Mempools, Snailchain, Preprocess, Daily Offchain Consensus, and on chain validation.

\subsection{Hybrid committee election}
\label{sec:election}

In \cite{pass2017hybrid}, $\mathrm{BFT}$ committee instances are switched after a fixed period of time (with the $\mathrm{snailchain}$ as a logical clock). Our $\mathrm{snailchain}$ is expected to produce one block every 10 minutes, so we've set a rotating frequency of 144 blocks. According to \cite{pass2017hybrid}, a new committee is formed simply by the miners of the latest $csize$ number of blocks inside $\mathrm{snailchain}$.

A naive implementation of \cite{pass2017hybrid} will be vulnerable to well known selfish mining attack strategies. The damage of selfish mining is magnified in the hybrid consensus setting, because power is more concentrated in the top few high hash nodes. If a selfish miner controlled more than 25\% of the blockchain's hash power, she could control more than 33\% of block production\cite{naya2015stub}\cite{eyal2013self}. Under the election procedure described in \cite{pass2017hybrid}, the selfish miner is likely to control over 1/3 of the BFT committee. If she happens to be adversarial, the $\mathrm{fastchain}$ will lose the liveness property.

There are two undesirable extremes that we need to find a balance in, 
\begin{itemize}
	\item Randomly select BFT members via VRF\cite{micali1999verifiable}. This is vulnerable against a sybil attack.
	\newline
	\item Any selection procedure where probability of being selected is proportional to hash power. The BFT committee is vulnerable against mining pools who are rich in hash power. 
\end{itemize} 

Our proposed solution are as follows. When an honest BFT node's chain reaches $\lambda$ in length, it will publish the unique miner IDs of every fruit in the chain as candidates (or, every miner ID who mined more than $\nu$ fruits). The new BFT committee is randomly selected from the candidates with uniform probability. The obvious sybil attack is not possible under this scheme, as you require a minimal level of PoW to become a candidate. Neither it will be easy for a large mining pool to achieve compromise the BFT committee. A fruit will be orders of magnitude easier to mine than a block. 

The uniformly distributed random number we use is similar to VRF~\cite{micali1999verifiable} where the seed is determined by the seed used in previous committee selection, as well as the proposed randomness
from recent $\mathsf{csize}$ blocks. Different from Algorand~\cite{gilad2017algorand}, here we don't count stake weights for this part of selection. Notice that the nodes that are chosen by random functions would have certain probability of not being online. This will cause our $\mathrm{fastchain}$ to lose chain quality. The user will need to manually flag that they are willing to participate in the committee election. 

\subsection{Secure communication channel}

The ability to establish a secure communication channel between committee nodes is central to the success of building a decentralized, high performant and secure network. There is a common misconception among the public that traditional PoW networks like Bitcoin and Ethereum are slow because of mining puzzles are too difficult to solve. On the contrary, the mining puzzle is nothing but a kill-time mechanism, so that the nodes have enough time to synchronize due to network delays. 

To make things worse, the aggregate network traffic required to achieve synchronization of all nodes, grows in polynomial order with respect to number of nodes in the network. As the network grows, its throughput will decay rapidly, making the development of its ecosystem unsustainable. The reason why PBFT based blockchains achieve far better transactions per second than PoW, is that the nodes required to reach consensus on a transaction is much fewer (typically 10 -- 30), and that stays constant regardless of the total number of nodes in the network, while we lose decentralization. 

Minerva aim to achieve high throughput by using a rotating BFT committee to process transactions, and PoW to determine who get elected to the committee. There lies a problem, how do BFT committee members communicate with eachother about their voting decisions? If the BFT committee use the gossip protocol of the PoW network, voting messages will need to traverse the entire PoW network before committee members hear about it. A real life experimental deployment of this network resulted in a throughput of 80 tps, a marginal improvement from Bitcoin cash. 

On the other hand, BFT committee members can make their communication channel known to the public, to establish direct peer-to-peer communication during the voting process. However, by (4) of our adversary model in section 4.1, we can expect this node to be DDoS'ed, and fast chain quality will rapidly deteriorate. Therefore, the central question we need to address is, how to protect the identities of rotating committee members while achieving direct peer-to-peer communication? 

Consider the following protocol, 

\begin{enumerate}
	\item If a PoW node mined over $\nu = 100$ fruits over the last 144 $\mathrm{snailchain}$ blocks, and is willing to participate in PBFT, then it automatically become a member of the candidate committee. 
	
	\item Hash data derived from fruits mined by node $i$ is used to an interval $I(i) \subset [0,1]$. Another random number, $\Gamma$, also generated from $\mathrm{snailchain}$ historical hash data (far enough so all nodes have synchronization). If $\Gamma \in I(i)$, then node $i$ has been elected to the committee. The value of $\Gamma$ should be the same across all honest nodes to achieve consensus. 
	
	\item Repeat (2) until enough committee members have been generated. 
	
	\item Let $\Gamma_1 , ..., \Gamma_{csize}$ be the generated uniform random numbers. Elected nodes can compute locally from $\mathrm{snailchain}$ data, the addresses of other members of the next committee. Suppose these are $\mathrm{add}_1, ... , \mathrm{add}_{csize}$. 
	
	\item Suppose node $\mathrm{add}_i$ is a new committee member. She will broadcast her IP address (or other data necessary to establish a direct communication channel), encrypted with the public key of $\mathrm{add}_1, ... , \mathrm{add}_{csize}$, using the gossip protocol. 
	
	\item When another committee member $\mathrm{add}_j$ hear this message, she will decrypt it with her private key, and find $csize - 1$ lines of gibberish and one line containing the IP address of node $\mathrm{add}_i$, while non-committee members will be unable to decrypt this message.  
\end{enumerate}

This protocol allow each committee member to locally build a table of IP addresses of other committee members, which will allow high throughput communication while hiding their IP address from the public. A slight drawback of this procedure is that if an adversary happens to be a committee member, she can broadcast the entire list of committee IP addresses to the general public. All of the committee nodes can be DDoS'ed in a short period of time, and the mildly adaptive adversary assumption is breached. 

Therefore, we make the following adjustment. Assume there are $csize$ members in the committee and we create a private gossip network, with each node gossiping to $gsize$ other nodes. A typical value of these parameters could be $csize = 31$ and $gsize = 4$. 

\begin{enumerate}
	\item We generate a matrix $A$ from historical hash data, such that $A_{i,j} \in \{ 0,1 \}$, and columns and rows of $A$ sum up to $gsize$. The matrix $A$ should be the same across all honest nodes to achieve consensus.
	
	\item We give label $i$ to the committee member who is elected by the random number $\Gamma_i$. 
	
	\item Committee member $i$ will only encrypt her IP address using the public key of $j$, provided if $A_{ij} = 1$.  
\end{enumerate}

We have just established a private gossip network of $csize$ nodes, each gossiping to $gsize$ other nodes. If an adversarial node made it to the BFT committee, the biggest damage is for the BFT committee to lose $gsize + 1$ nodes. Setting $gsize$ big could improve the blockchain's throughput, while a smaller $gsize$ improves the chain's security. By experiment we found that setting $gsize < \frac{1}{6} csize$ generally gives a good balance. This allowed us to achieve over 2000 transactions per second in beta-net live deployment setting. 

If $gsize$ is too small, we could run the risk of having an eclipse attack, particularly when accompanied by a bad choice of $A$. Roughly speaking, if $A$ is very close to a direct sum of lower dimensional matrices, the resulting graph will start to have isolated regions, making some nodes a more critical connector than others. The algorithm to consistently generate robust choices of $A$, relies on the representation theory of symmetric groups, which is will be explained in a forthcoming paper.

\subsection{Application Specific Design}

Our consensus design is aware of application specific requirements and tailors for them, under the conditions that the consistency, liveness
and security properties are not compromised.

\subsubsection{Physical Timing Restriction}

Conventional consensus design by default allows miners / committee members / leaders to re-order transactions within a small timing window.
This raises a problem for some decentralized applications such as commercial exchanges where the trading fairness requires the timing order between
transactions to be carefully preserved, or otherwise malicious (or, even normal rational) participants will have the incentive to re-order transactions,
or even insert its own transactions, to gain extra profits. And this incentive will be magnified under high throughputs.

And what is even worse, is that such malicious re-ordering is impossible to distinguish because naturally network latency will cause re-ordering and such latencies
can only be observed by the receiver itself and therefore it has the final evidence of numbers regarding network latency.

To support decentralized advertisement exchanges, we try to reduce such problems by incoporating one more restriction called sticky timestamp.
More specifically, with a heuristic parameter $T_\Delta$, when proposing transactions, we require the client to put a physical timestamp $T_p$ inside
the metadata of the transaction, and this physical timestamp is signed together with the other parts of the transaction. Later when validators inside
$\mathsf{BFT}$ verify the transaction, it will do the following extra checks as shown in Algorithm~\ref{alg:timestamp}.

\begin{figure*}
\begin{algorithm}[H] 
\KwData{Input Transaction $\mathsf{TX}$}
\KwResult{A Boolean value that indicates whether the verification is passed}
$\mathsf{current\_time} \gets \mathsf{Time.Now()}$\;
\If{$|current\_time-\mathsf{TX.T_p}|>T_\Delta$}{
	return $\mathsf{false}$\; \tcp{if the time skew is too large, reject $\mathsf TX$.}
}
var $\mathsf{txn\_history=}$ new static dictionary of lists\;
\eIf{$\mathsf{txn\_history[TX.from]} == NULL$}{
	$\mathsf{txn\_history[TX.from]} == [TX]$\;
}{
	\eIf{$\mathsf{txn\_history[TX.from][-1].T_p}-\mathsf{TX.T_p} > 0$}{
		return $\mathsf{false}$\; \tcp{To make sure the transactions from the same node preserve timing order.}
	}{
		$\mathsf{txn\_history[TX.from].append(TX)}$\;
		return $\mathsf{true}$\;
	}
}
\caption{Extra Verification Regarding Physcal Timestamp}
\label{alg:timestamp}
\end{algorithm}
\caption{Pseudo-Code for Extra Verification}
\end{figure*}

At the stage of materializing logs inside $\mathsf{BFT}$, the leader will sort the transaction batch according to its physical timestamps and break ties
(though very unlikely) with the sequence number. Actually this step is not necessary because we can enforce the order later in the evaluation and verification.
But for simplicity, we put it here.

This set of modications give us several extra properties:

\begin{enumerate}
\item The order of transactions from any node $N_i$ is internally preserved according to their physical timestamps. Thus the sequence order of these transactions
is strictly enforced. This will get rid of the possibility of some malicious re-ordering that involves two transcations from the same node.
\item The order within a batch of transactions output by the $\mathsf{BFT}$ committee is strictly ordered by timestamps.
\item Nodes cannot manipulate fake physical timestamps because of the timing window restriction.
\end{enumerate}

One obvious disadvantage of this modificaion will be the reduction in terms of throughput due to aborting transactions when the parameter $T_\Delta$ is inappropriate
for the varying network latency. Another disadvantage is that, the $\mathsf{BFT}$ committee members are still allowed to lie about their local time and reject certain
transactions. However, committee members can reject certain transactions anyway. But honest nodes could potentially reject ignorant transactions because of their
unsynchronzied clocks. This issue can be reduced by adding restrictions on the eligibilities of the $\mathsf{BFT}$ committee. Later we will see that to get into
the committee, the nodes should present evidence of synchronized clocks.

\section{Computation and Data Sharding, and Speculative Transaction Execution}
In this section we introduce our sharding scheme.

An important modification over the original Hybrid Consensus is that we add computation and data sharding support for it. And even more, first of its kind,
we design a speculative transaction processing system over shards. The idea is clear. In Hybrid Consensus, the $\mathsf{DailyBFT}$ instances are indexed into
a determininstic sequence $\mathsf{DailyBFT[1\dots R]}$. We allow multiple sequences of $\mathsf{DailyBFT}$ instances to exist at the same time. To be precise,
we denote the $t$-th $\mathsf{DailyBFT}$ sequence by shard $S_t$. For simplicity, we fix the number of shards as $C$. Each $\mathsf{DailyBFT}$ is a normal shard.
Besides $C$ normal shards, we have a primary shard $S_p$ composed of $\mathsf{csize}$ nodes. The job of the primary shard is to finalize the ordering of the
output of normal shards as well as implementing the coordinator in distributed transaction processing systems. And the normal shards, instead of directly connecting
with Hybrid Consensus component, submit logs to the primary shard, which in turn talks to Hybrid Consensus.

We don't allow any two shards (either normal or primary) to share common nodes, which can be enforced in the committee selection procedure. The election of
multiple shards is similar to the election procedure described in Section~\ref{sec:election}.

We partition the state data (in terms of account range) uniformly into $C$ shards. This will make sure that every query to the corresponding shard will return
a consistent state. Since we are going to include meta data for each data unit, we split data into units of data sectors and assign each data sector with an
address. We have a mapping from data position to data sector address. For simplicity, from now on, we only discuss at the level of data sectors. Each
data sector $\mathsf{DS[addr]}$ has metadata of $\mathsf{rts, wts, readers, writers}$.

We assume the partition principle is public and given the address $\mathsf{addr}$ we can get its host shard by calling the function $\mathsf{host(addr)}$.

Notice that if we treat every normal shard (when the number of adversaries is not large) as a distributed processing unit, we can incorporate the design of
logical timestamps \cite{yu2016tictoc} in distributed transaction processing systems \cite{mahmoud2014maat}, which will empower the processing of transactions.
Here we utilized a simplified version of MaaT where we don't do auto-adjustment of other transaction's timestamps.

For normal shards, it acts exactly as described in $\mathsf{DailyBFT}$ except the following changes to make it compatible for parallel speculative execution.

\begin{figure*}
\begin{algorithm}[H]
\textbf{On BecomeShard:}\\
\myin Initialize all the state data sectors: $\mathsf{lastReaderTS = -1, lastWriterTS = -1, readers=[], writers=[]}$

\textbf{With transaction $\mathsf{TX}$ on shard $S_i$:\\
On Initialization:}\\
\hspace{0.1in} $\mathsf{TX.lowerBound = 0}$\;
\hspace{0.1in} $\mathsf{TX.upperBound = +\infty}$\;
\hspace{0.1in} $\mathsf{TX.state = RUNNING}$\;
\hspace{0.1in} $\mathsf{TX.before = []}$\;
\hspace{0.1in} $\mathsf{TX.after = []}$\;
\hspace{0.1in} $\mathsf{TX.ID = rand}$\;

\textbf{On Read Address($\mathsf{addr}$):}\\
\eIf{$\mathsf{host(addr)}==S_i$}{
	Send $\mathsf{readRemote(addr)}$ to itself\;
}{
	Broadcast $\mathsf{readRemote(addr, TX.id)}$ to $\mathsf{host(addr)}$\;
	Async wait for $2f+1$ valid signed replies within timeout $T_o$\;
	Abort $\mathsf{TX}$ when the timeout ticks\;
}
Let $\mathsf{val, wts, IDs}$ be the majority reply\;
$\mathsf{TX.before.append(IDs)}$\;
$\mathsf{TX.lowerBound= max(TX.lowerBound, wts)}$\;
return $\mathsf{val}$\;

\textbf{On Write Address($\mathsf{addr}$):}\\
\eIf{$\mathsf{host(addr)}==S_i$}{
	Send $\mathsf{writeRemote(addr)}$ to itself\;
}{
	Broadcast $\mathsf{writeRemote(addr, TX.id)}$ to $\mathsf{host(addr)}$\;
	Async wait for $2f+1$ valid signed replies within timeout $T_o$\;
	Abort $\mathsf{TX}$ when the timeout ticks.
}
Let $\mathsf{rts, IDs}$ be the majority reply\;
$\mathsf{TX.after.append(IDs)}$
$\mathsf{TX.lowerBound= max(TX.lowerBound, rts)}$\;
return\;

\textbf{On Finish Execution:}
\For{every $\mathsf{TX'} in TX.before$}{
	TX.lowerBound = max(TX.lowerBound, TX'.upperBound)\;
}
\For{every $\mathsf{TX'} in TX.after$}{
	TX.upperBound = min(TX.upperBound, TX'.lowerBound)\;
}
\If{TX.lowerBound > TX.upperBound}{
	Abort TX\;
}
Broadcast $Precommit(TX.ID, \lfloor \frac{TX.lowerBound + TX.upperBound}{2}\rfloor)$ to all the previous remote shards  which $TX$ has accessed\;
\tcp{If TX.upperBound = $\infty$, we can set an arbitrary number larger than $TX.lowerBound$.}

\textbf{On receive $\mathsf{readRemote(addr, ID)}$:}\\
\eIf{$\mathsf{host(addr)}==S_i$}{
	$\mathsf{DS[addr].readers.append(ID)}$\;
	return $\mathsf{DS[addr].value, DS[addr].wts, DS[addr].writers}$\;
}{
	Ignore
}

\textbf{On receive $\mathsf{writeRemote(addr, ID)}$:}\\
\eIf{$\mathsf{host(addr)}==S_i$}{
	$\mathsf{DS[addr].writers.append(ID)}$\;
	Write to a local copy\;
	return $\mathsf{DS[addr].rts, DS[addr].readers}$\;
}{
	Ignore
}

\caption{Sharding and Speculative Transaction Processing}
\label{alg:shard}
\end{algorithm}
\caption{Pseudo-Code for Sharding and Speculative Transaction Processing}
\end{figure*}

\begin{figure*}
\begin{algorithm}[H]
\textbf{On receive $\mathsf{Precommit(ID, cts)}$:}\\
Look up TX by ID\;
\If{Found and $\mathsf{cts}$ not in $\mathsf{[TX.lowerBound, TX.upperBound]}$}{
	Broadcast $Abort(ID)$ to the sender's shard.\;
}
TX.lowerBound = TX.upperBound = cts\;
For every data sector $DS[addr]$ $\mathsf{TX}$ reads, set $DS[addr].rts = max(DS[addr].rts, cts)$\;
For every data sector $DS[addr]$ $\mathsf{TX}$ writes, set $DS[addr].wts = max(DS[addr].wts, cts)$\;
Broadcast $Commit(ID, batchCounter)$ to the sender's shard.\;
\tcp{batchCounter is a number which increases by 1 whenever the shard submit a batch of log to the primary shard.}

\textbf{On receive $2f+1$ $\mathsf{Commit(ID, batchCounter)}$ from each remote shards which $TX$ has accessed:}\\
TX.lowerBound = TX.upperBound = cts\;
For every data sector $DS[addr]$ $\mathsf{TX}$ reads, set $DS[addr].rts = max(DS[addr].rts, cts)$\;
For every data sector $DS[addr]$ $\mathsf{TX}$ writes, set $DS[addr].wts = max(DS[addr].wts, cts)$\;
Mark $TX$ committed\;
Let $TX.metadata = [ShardID, batchCounter]$\;

\textbf{On output log}\\
Sort $TX$'s based on their $cts$. Break ties by physical timestamp.

\caption{Sharding and Speculative Transaction Processing (cont.)}
\end{algorithm}
\end{figure*}

For the primary shard, it collects output from all the normal shards. Notice that, the data dependency of transactions can be easily inferred by their metadata.
And a fact is that, if a transaction visits multiple remote shards, it will leave traces in all the shards involved. When a normal shard submit logs to the primary
shard, it will also write to the snailchain.

When the primary shard receives (or fetchs from the snailchain) a batch of txns from a shard, it will check if it has received from all the shards transactions
within this batch. If after certain timeout it has not received transactions from a particular batch, it means that batch has failed. In this case, a whole committee
switch will be triggered at the next day starting point. After receiving all the shards' logs, the primary shard sorts the transactions based on their commit
timestamps (if some transaction has earlier batch number, it will be considered as the first key in the sorting, however, if its physical timestamp violates
the timestamps from many shards, we decide that batch as invalid and all the transactions inside that batch are aborted). After sorting, the primary shard
filters all the transactions and keeps a longest non-decreasing sequence in terms of physical timestamps. Out the log to the Hybrid Consensus component as that
day's log.

There are still many optimisation spaces. One certain con is that the confirmation time in this design is not instant.


\section{Smart Contracts in Virtual Machines}

\subsection{Design Rationale}

Of all the reasons to have an Ethereum Virtual Machine (EVM)~\cite{gavinethereum}, one of aims is to meter the usage with a transaction fee
in a Proof of Work model. Since ours is a hybrid model, we'll take the liberty of exploring this design space a little bit further. Let us
consider the possibility of a hybrid cloud ecosystem.

A basic problem people have faced is the kind of crude mathematical notations followed in Ethereum's Yellow Paper~\cite{gavinethereum}.
We therefore hope to follow something like KEVM jellopaper~\cite{kevmjello} to list out the EVM and TVM (described in~\ref{sec:tvm}) specifications.
And in future, we hope to maintain our own specifications through Truechain's github account (https://github.com/truechain).

\subsubsection{What about containers instead of VMs?}

One of the blockchain frameworks out there that come as close to this idea as possible, is Hyperledger's Fabric framework~\cite{fabricpaper}.
If one sets out to convert Fabric's permissioned nature into permissionless, one of the foremost challenges
would be to solve the chaincode issue. What this means is while it's possible to keep a chaincode/smart contract in a single container,
that is not a scalable model for a public chain. Having such a model for public chain means having to run several thousand containers,
per se, several thousand smart contracts on a single node (because each node maintains a copy).

\smallskip
There have been attempts from the community on being able to run a certain maximum containers per node. The limit currently is 100 pods
per node, per se, approximately 250 containers per node, as illustrated in Kubernetes container orchestration platform~\cite{kubelimit}
and Red Hat's Openshift Container Platform 3.9's Cluster Limits~\cite{ocpclusterlimit}. Even with latest storage techniques like
brick multiplexing~\cite{brickmul}, the max possible value (say $\mathsf{MAX\_CONTR}$) of containers could not possibly reach
(at least right now) 1000. This issue could further be looked up in the discussions on kubernetes issues
github page~\cite{k8scale} around workload-specific limits that usually determine the maximum pods per node.
People who wish to scale containers usually prefer horizontal scaling rather than a vertical scaleup~\cite{cncfscaleout, kubecommgoals},
as the latter significantly increases complexity of design decisions. And there's no one-size-fits-them-all rule for a cluster scale
configuration as that entirely depends on the workload, which being more in our case due to its decentralized nature, isn't very
convincing for taking a step towards scaling this. At this point, it becomes more of an innovation problem than a simple technical
specification search. Ethereum currently has $\mathsf{ > 1000 }$ smart contracts deployed. Therefore this would be nothing but a crude
attempt at optimizing the container ecosystem's design space.

Now let us expand a bit on the container scenario. Given the above crisis, a possible solution is to use container in a
serverless architecture. But consider a scenario where $\mathsf{> 2000}$ contracts are online and the concurrent requests,
i.e., invocation calls to chaincode (a moving window) at a time exceed $\mathsf{MAX\_CONTR}$ value, we then face the same
problem all over again. Therefore, it is only advisable to add a throttling rate limit on the max concurrent requests.
This severly limits the Transactions Per Second from the consensus, by design. Engineering should not be a bottleneck to what could
be achievable alternatively. Therefore, we choose to stick to EVM design, although a bit modified for our purpose.

\subsection{Truechain Virtual Machine (TVM)}
\label{sec:tvm}

A typical example in this space would be that of the Ethereum Virtual Machine (EVM) ~\cite{gavinethereum}, which tries to follow
total determinism, is completely optimized and is as simple as it gets, to make incentivization a simple step to calculate.
It also supports various features like off-stack storage of memory, contract delegation and inter-invocation value storage.

We would reuse the EVM specifications for the snailchain, but add a new specification for TVM in the next version of this Yellow Paper,
after careful consideration of the design rationale similar to EVM, deriving the stack based architecture utilizing the Keccak-256
hashing technique and the Elliptic-curve cryptography (ECC) approach.

The Truechain infrastructure will utilize a combination of EVM and another EVM-like bytecode execution platform for launching smart contracts.
We choose to use run VM only on $\mathrm{fastchain}$, embedded within each full node, so they could manage invocation calls on per-need basis.

The TVM backs the DailyBFT powered chains, which interact with the following components:
\begin{itemize}
  \item re-using some of the concepts from tendermint, like the ABCI (Application BlockChain Interface) which offers
  an abstraction level as means to enable a consensus engine running in one process to manage an application state running in another.
  \item A different consensus engine pertaining to dailyBFT chain,
  \item A permissioned Ethereum Virtual Machine
  \item An RPC gateway, which guarantees (in a partially asynchronous network) transaction finality
\end{itemize}


\section{Incentive design and gas fee}
The Proof of work protocol have a proven track record of attracting computational resources at an unprecedented rate. While existing PoW networks such as bitcoin and ethereum have been successful in their own right, the computational resources they attracted have been nothing more than very powerful hash calculators. They cost a lot of electricity to run, and produce nothing useful. 

In this section we will present a concept of compensation infrastructure in order to balance the workload of $\mathsf{BFT}$ committee members and non-member full nodes. We invented a new incentive design for PoW, where participating resources can be redirected to do useful things, such as scaling transactions per second (referred to as ``TPS" from hereon), and providing on-chain data storage. 

Ethereum gas price is determined by a spot market with no possibility of arbitrage, similar to that of electricity spot market studied in \cite{schmidt08}. We consider this market to be incomplete, and therefore fundamental theorem of asset pricing does not apply\cite{delbaen94}. Thus, the underlying gas price will follow a shot-noise process, that is known for its high volatility. We introduce a ``gas market place" where gas will be traded as futures, and this market is complete in the infinitesimal limit. This is expected to significantly reduce gas price volatility compared to Ethereum.

The following subsections will talking about each component of the incentive design in detail. 

\subsection{ASIC resistance}

Truechain's mining algorithm will be fundamentally ASIC resistant. We will define what this means, and why Truehash can achieve fundamental ASIC resistance in a forthcoming paper. Here, we will briefly outline what we have done. ASIC's are far better hash calculators than any general purpose computer, but they can only one procedure and cannot be reprogrammed. Therefore, if there is an automated procedure that would switch the mining algorithm once every few months (12000 $\mathrm{snailchain}$ blocks), then it would not be profitable for anyone to build them. 

We came up with the following conditions for fundamental ASIC resistance, 

\begin{enumerate}
	\item There exist a large pool of potential mining algorithms (e.g. 2048!) that is unfeasible to hard code each of them to an ASIC chip. 
	
	\item The mining algorithm is capable of automatically switching without having human interference, such as a hard fork. 
	
	\item The correctness of the new mining algorithm must be provable (so all nodes can come to consensus), and the process must be verifiably unpredictable (so its impossible to secretly build ASIC miners before the switch).  
\end{enumerate}

The classical PoW mining basically repeated calculate $\mathrm{hash}(v(nonce))$, where $v(nonce)$ is a function that takes the mining $nonce$, block header and pads it to the correct dimension. We compute instead, the modified hash function, $\mathrm{hash}(\rho (g)*v(nonce))$. Here, $G$ is a large group (e.g. $S_{2048}$) and $g \in G$, $\rho :G \rightarrow V$ is a homomorphism from $G$ to a vector space $V$, so that the matrix multiplication $\rho (g)*v(nonce)$ is valid on $V$. 

As we can see, by simply replacing $\rho (g)$ to $\rho (g’)$, for a different element $g' \in G$, we get a completely different hash algorithm. Since $G = S_{2048}$ is a large group, $|G| = 2048!$, the space of potential hash algorithms to choose from is 2048!. 

Finally, we can generate the new group element $g'$ from the hash data of $\mathrm{snailchain}$, through application of group representation theory. The new mining algorithm is therefore provable and unpredictable.

\subsection{Gas fee and sharding}

Gas price is traded in a futures market, where the futures contract is manifested by a smart contract. Specifically, the contract will be executed as follows. 

\begin{itemize}
	\item Party A agree to pay party B $xxx$ TRUE, while party B promises to execute party A's smart contract, between time $T_0$ and $T_1$, that cost exactly 1 gas to run.
	
	\item Party B will contribute $xxx$ TRUE to a pool corresponding to the committee C that executed party A's smart contract. This is called the gas pool. 
	
	\item Members of C will receive an equal share of the pool, and return an average cost per gas $\mu$ for the pool.
	
	\item If B contributed less than $\mu$, she must make up for the difference by paying another party who contributed more than $\mu$. If B contributed more than $\mu$, she will receive the difference from another party. 
	
\end{itemize}     

Under this scheme, liquidity providers are rewarded when they correctly anticipate network stress, and hence creating a complete market in the infinitesimal limit. Price volatility are absorbed by the averaging mechanism in the gas pool making the price itself a good indicator of network stress. 

Our intention to ensure gas price is traded roughly within a predetermined interval. Hence, if the moving average price sustain above a certain threshold, a new PBFT committee is spawned through a quantum appearance process. On the other hand, if the moving average price sustain below a certain threshold, an existing PBFT committee will not be given a successor after it finished serving its term.

The proportion of mining reward will be distributed as follows. Let $n$ be the number of PBFT committee running at a certain instance, and $\alpha > 1$. Proportion of mining reward going to PBFT nodes is equal to $\frac{n}{\alpha + n}$, and PoW nodes $\frac{\alpha}{\alpha+n}$. This is to reflect that in later stages of the chain, new nodes are incentivized to contribute to the blockchain's overall TPS, hence ensuring continued scalability of the chain. The parameter $\alpha$ represent the number of PBFT committees when mining reward is divided 50-50.

\section{Data storage}

Truechain aim to achieve 10,000 TPS on each shard, and number of shards is designed to grow linearly with respect to the volume transaction demand (roughly correlates to number of nodes). The bitcoin network has generated roughly 170 gb of transaction history data over the course of 10 years at 10 TPS. Going from 10 to 10,000, the same volume of data is expected to be generated every 3 days. At 100 shards, or 1 million TPS, we can expect it to generate every 45 minutes. Hence, having every node to store the entire transaction history like bitcoin, is no longer feasible in a high TPS public chain. 

A number of solutions have been proposed such as having a check point once every few hours/days, where every node is only required to store transaction history from previous $n$ check points. But where do we store rest of the transaction history, as nobody is incentivized to store that. 

Our solution is to seamlessly merge transaction processing with the storage capability of an IPFS in a unified incentive infrastructure. This will provide a solution to store transaction history, and allow a plethora of sophisticated applications to be running completely decentralized on the Truechain architecture. Data storage on Truechain will be possible in three levels, 

\begin{itemize}
	\item Level 1: Stored on every PoW node like bitcoin and ethereum. It is the most permanent way of storage, alas also the least efficient. It's predominately designed to store the block's Merkel root, and other information of high value density. Users will pay a gas fee to PoW miners. 
	
	\item Level 2: There will be an IPFS-like file system where a limited copy of the data is distributed to storage nodes throughout the chain. This is designed for storing the bulk of main net's historical transactions and data-heavy decentralized applications. Users will pay miners a fee for both storage and retrieval. 
\end{itemize}

\section{Future Direction}

Even after optimizations to the original Hybrid Consensus, we acknowledge various optimizations possible on top of what was
proposed in this paper. There are following possibilities:

\begin{itemize}
  \item Improving timestamp synchronization for all nodes, with no dependency on centralized NTP servers.
  \item Detailed incentivization techniques for compensation infrastructure, so heavy infrastructure investors don't suffer from 'left-out', 'at a loss' problem
  \item Sharding techniques with replica creation to minimize the transaction set rejection from the BFT committee.
  \item Addition of zero knowledge proof techniques for privacy.
  \item Hybrid infrastructure of EVM, TVM and Linux container ecosystem.
  \item Sections for Virtual Machine Specification, Binary Data Encoding Method, Signing Transactions, Fee schedule and Ethash alternative.
\end{itemize}

\section{Conclusions}

We have formally defined Hybrid Consensus protocol and its implementation along with plausible speculations in the original
proposal. In this draft, we have introduced various new concepts some of which we will detail in the next version very soon.
We recommend people to choose ASIC resistant hardware for deployment of the PoW only versus full nodes,
although more details on hardware shall follow soon.

\begin{itemize}
  \item A permissioned BFT chain that runs on a few nodes in the permissionless PoW based network.
  \item The BFT committee is a rotating one, preventing corruption in a timely manner
  \item The BFT committee is responsible for transaction validation, and the PoW nodes are only responsible for choosing/electing the committee members according to some rules we've derived and re-defined.
  \item The new permissioned VM, we've surmised, could be inspired from the EVM, but with different block states and transaction execution flows
  \item The contemporary permissionless EVM in the PoW chain co-exists with this new permissioned VM (which we call Truechain Virtual Machine - TVM)
  \item The TVM would be the one to validate any transactions related to consensus, while the traditional EVM would need to be re-worked to not really mine for consensus, but for election of BFT using Variable Day length puzzle.
  \item The incentivation model needs to be re-worked such that it is based off of TVM, and we still reward the miners in PoW chain.
  \item We invented a new market mechanism, satisfying the assumptions of fundamental theorem of asset pricing, for how gas should be traded. 
  \item We would eventually support sharding for the BFT committee nodes, for scalability.
  \item We address the storage issue for high TPS public chains, and introduced a method that seamlessly merge transaction process with decentralized data storage. 
  \item A compensation infrastructure, which accounts for node configuration non-uniformity (different CPU/memory/network bandwidth in the node pool), would eventually be a part of the consensus, thus speeding up transactions.
  \item The smart contracts execution would thus only happen in TVM (BFT node).
\end{itemize}

\section{Acknowledgements}

We owe a great deal of appreciation and are thankful, to the following folks for their untiring work towards pushing the protocols
for a decentralized sovereignty, for their design rationale and implementations which served as a solid reference architecture
in our proposal above. These folks and their legacies are as mentioned below:

\begin{itemize}
  \item Rafael Pass, Miguel Castro, Satoshi Nakamoto, Vitalik Buterin, Gavin Wood, Ethan Buchman, Andrew Miller et al
  for their untiring work, contributions and continuous improvisations while spearheading the glamorous Improvement Proposals forums in
  addition to the active participation through Reddit, Mailing lists, chat forums, white and Yellow Papers, and rest of the mediums alike.
  \item CNCF and Kubernetes communities for their inspiring ventures into hybrid cloud computing.
\end{itemize}


\end{multicols}

\bibliography{ref}
\bibliographystyle{abbrv}

\appendix

\section{Terminology} \label{ch:Terminology}

\begin{description}
\item[TrueChain Virtual Machine (TVM)] In contrast to EVM which handles incentivization and rotating committee selection,
  a TVM is based on similar design principles but carries out actual consensus and voting based off of PBFT based Hybrid Consensus.
\end{description}


\end{document}